\documentclass[aps,prl,twocolumn,showpacs,floats]{revtex4}
\usepackage{graphicx}%
\usepackage{epsfig}
\usepackage{amsmath}%
\usepackage{amssymb}%

\def\beq{\begin{equation}}
\def\eeq{\end{equation}}
\def\e{\epsilon}
\def\half{{\ss 1\over 2}}
\def\ksi{\xi}
\def\ss{\scriptstyle}

\def\prl{{\sl Phys. Rev. Lett.}\ }
\begin{document}
\bibliographystyle{prsty}
\title{\Large\bf Quantum Hall Criticality, Superconductor-Insulator
 Transition and Quantum Percolation}
\author{Yonatan Dubi$^1$, Yigal Meir$^{1,2}$ and Yshai Avishai$^{1,2}$}
\affiliation{
$^{1}$ Physics Department, Ben-Gurion University, Beer Sheva 84105, Israel\\
$^{2}$ The Ilse Katz Center for Meso- and Nano-scale Science and
Technology, Ben-Gurion University, Beer Sheva 84105, Israel }
\date{\today}

\begin{abstract}
A model consisting of a mixture of superconducting and quantum
links is proposed to describe the integer quantum Hall transition.
The quantum links correspond to tunneling of electrons between
trajectories trapped in adjacent potential valleys, while the
superconducting links correspond to merging of these trajectories
once the Fermi energy crosses the saddle point energy separating
the
 two valleys. The quantum Hall transition in this model corresponds to
 percolation of the superconducting links. Numerical calculations and
  scaling analysis using two different approaches
 yield the critical exponent $\nu \approx 2.4$ and a two-peak conductance
    distribution at the
critical point. The role of quantum coherence is discussed, and an explanation
  of experimental observations claiming different universality class
  for the quantum Hall transition is suggested. The model suggests that the
  critical behavior of the superconductor-insulator transition from the
  insulating side is in the same universality class as the quantum Hall transition.
\end{abstract}
\maketitle

 The integer quantum Hall (QH) effect is one of the most studied examples
 of a second order quantum phase transition \cite{QHEreview}. As such it is
  characterized by a diverging length scale $\xi$, describing the
  typical decay of a wavefunction, or of the conductance.
 The physics  describing the transition seems well understood, and most numerical
  investigations in various models yield $\nu\simeq2.4$ \cite{huckenstein},
   where $\nu$
   describes the divergence of $\xi$ at the critical point.
  Experiments, however, disagree on the critical behavior.
  While some indeed yield an exponent close to
  the theoretical prediction \cite{exp2.4}, others yield an exponent
  $\nu\simeq1.3$, close to the classical percolation critical exponent,
  $\nu_p=4/3$ \cite{shahar}.
  In fact, the role of percolation in the QH transition
   has been experimentally
   demonstrated \cite{QHEperc}. Yet other experiments
   claim that there is no critical behavior \cite{nocritical}.

 Similar lack of clarity exists with respect to the critical exponent describing
 the two-dimensional superconducting-insulator transition (SIT),
 which can be induced either by a continuous change
 of film thickness, or by a magnetic field. In both cases some
 experiments report an exponent $\nu$ close to $1.3$ on the insulating side
 \cite{sc1.3},
  while others yield $\nu\simeq2.8$ \cite{sc2.8}. On the superconducting side
  only $\nu\simeq1.4$ has been reported. Again, similarly to the QH case,
  some experiments give evidence of
  no critical behavior \cite{yazdani}. The relevance of the percolation of
  superconducting areas to the SIT has been pointed out long time
  ago \cite{deutscher}, and has been demonstrated experimentally both in
  granular systems \cite{gerber} and in amorphous systems \cite{ovadyahu}.
  More recent scanning tunneling measurements \cite{millo} have
  directly demonstrated the separation of these systems into superconducting
  areas and metallic or insulating ones.

   In this work we develop a generic model to describe the QH transition.
   This model, which is directly related to the SIT, and to quantum percolation,
   clarifies the interplay of classical percolation and quantum
   tunneling and interference in describing the critical behavior
   of the QH transition. In addition to providing an estimate of the
    critical exponent and the fixed-point conductance distribution at the
   critical point, it enables us to investigate the role of tunneling
 and interference on the critical behavior, suggesting an explanation for
the appearance of the classical critical exponent in the QH
transition in some experiments \cite{shahar}. Moreover, the model suggests
 that the SIT, on the insulating side,
 belongs to the same universality class as the QH transition, explaining
 the closeness of the numerical and experimental estimates of the critical exponents of both transitions.

\par  The model proposed here is based on the following picture. In strong magnetic fields
electrons perform small oscillations around equipotential lines.
At low energies, their trajectories are trapped around potential
valleys, with quantum tunneling occurring between adjacent
valleys. We associate each such potential valley with a discrete
lattice site, where nearest neighbors are connected by links that
correspond to random tunneling between these trapped
trajectories (see Fig.~1). As the Fermi energy $\e_F$ rises and crosses
the saddle point energy separating two neighboring valleys, the
two isolated trajectories coalesce, the electron
 can freely move from one valley to its neighbor and
 the link connecting them becomes perfect or "superconducting" (SC). The QH
  transition occurs when one trajectory spans the whole system, which
 in this lattice model corresponds to percolation of the SC
 links. The critical behavior, however, will be determined by the
 quantum mechanical processes involved in transport of an electron from one
 side of the system to the other, namely tunneling and interference.

\par \noindent For simplicity, the valleys are  mapped onto sites on a square
lattice. The saddle-points, which are mapped onto links in the
lattice, have random energies $\e_i$, sampled from some
distribution $G(\e_i)$.
 For a given Fermi energy, the transmission through the link for $\e_F<\e_i$ is
given by
\begin{equation}
   T(\e_F)=\exp\left[-\alpha(\e_i-\e_F)\right],
  \label{TE}
\end{equation}
describing tunneling through a parabolic barrier, with $\alpha$ a
positive constant. For $\e_F>\e_i$, on the other hand, the transmission through the
 link is perfect (i.e. $T(\e_F)=1$) and the link is considered to be SC.
 At low Fermi energy, very few SC links exist, and their concentration $p(\e_F)$,
given by $
 p(\e_F)=\int_{-\infty}^{\e_F} G(\e) d\e, $
   increases with $\e_F$. When $p(\e_F)=p_c=\half$, the critical
    threshold for the square lattice, the SC links percolate through the
     system and the localization length diverges, signaling the QH
     transition.
      We will be interested in the critical behavior near
   this phase transition point \cite{twosides}.


\begin{figure}[!h]
\centering
\includegraphics[width=6.2truecm]{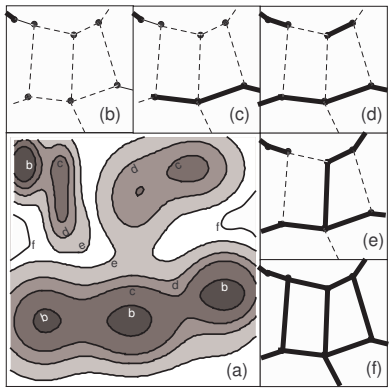}
\caption{\small Mapping of the quantum Hall system onto a discrete
model. (a) Equipotential lines for five different Fermi energies.
At low Fermi energies electron trajectories are confined to
potential valleys. Each such valley is denoted by a point in the
discrete model (b)-(f). In order to move from one such valley to
another, the electron has to tunnel through a saddle point --
dotted links in (b)-(f). As the Fermi energy rises above the
saddle point energy
 these valleys are joined and electron can move without
dissipation from one valley to another -- solid links in (b)-(f).
Figures (b)-(f) correspond to a system with Fermi energy up to the equipotential line
pointed by the appropriate letter in (a). Note that in (d) the solid links percolate,
meaning that at this
Fermi energy an electron can traverse the whole system on an
equipotential line, corresponding to the quantum Hall transition.
}
\end{figure}

Two computational approaches for solving this quantum mechanical
problem are presented. The first one employs a scattering matrix
formalism. Each link in the lattice carries two edge states
from neighboring valleys, moving in opposite directions (see
Fig.~2a). The scattering matrix for each link is characterized by
complex transmission and reflection amplitudes associated with
electron tunneling between adjacent valleys. The tunneling
probability is determined as follows: for each link, a
saddle-point energy $\e_i$ is randomly chosen from a uniform
distribution $G(\e_i)$ on $[-\half,\half]$ (so that the critical
energy is $\e_c=0$, corresponding to $p_c=\half$). If $\e_F>\e_i$ then $T=1$, otherwise  the
transmission probability through the link is given by (\ref{TE}).
The allowed phases of the matrix elements are chosen randomly from
a uniform distribution between $0$ and $2\pi$.

For perfect links ($T=1$), the electron follows these edge states
from one valley to another, according to their chirality. It is
easy to see (Fig.~2b) that when these percolate, we have two edge
states propagating through the system in opposite directions,
without scattering between left going and right going channels. Consequently, 
the percolation point corresponds to the quantum Hall transition.

\begin{figure}[!h]
\centering
\includegraphics[width=8.4truecm]{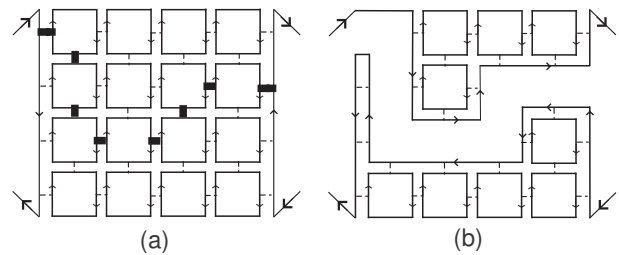}
\vskip -0.3truecm
\caption{\small The scattering matrix approach: each
link carries two counter propagating edge modes (a). A non zero
transmission (broken lines) allows electrons to tunnel between
adjacent sites (potential valleys). When the transmission is unity
(bold lines in (a)), these two valleys merge, and an edge
state can freely propagate from
 one to another. A percolation of these perfect transmission links (b)
  correspond to an edge state propagating through
the system without backscattering. } \vskip -0.4 truecm
\end{figure}

We evaluate the transmission $T(L)=\exp(\langle \log T \rangle)$,
through a system of the geometry depicted in Fig.~2, of linear
sizes $L=10,15,\ldots,50$, averaged over $5000$ realizations,
  for different values
   of the Fermi energy near the critical point.
 Fig.~3 displays the raw data. In (a) we plot  the dependence
   of the transmission on Fermi energy (in terms of probability of SC links)
   for different sizes. The existence of a critical point, where
   the curves cross and the transmission is independent of length,
   is evident.
   In (b) we show the dependence of the transmission on length for different concentrations. The inset depicts
   the log of the transmission as
    a function of length. The straight lines demonstrate that indeed,
    $T(L)\propto \exp(-L/\ksi(\e_F))$.
   All these curves coalesce (Fig.~4(a)), after scaling the system length the
   energy-dependent localization length $\ksi(\e_F)$. $\ksi(\e_F)$ is found to diverge
   at the critical point with an exponent $2.34\pm0.1$, in agreement with previous
   numerical approaches. The inset of Fig.~4(a) displays the distribution of
   transmissions at the critical point. The double peak structure agrees with previous
   numerical calculations for the integer QHE \cite{distribution}.

\begin{figure}[!h]
\centering
\includegraphics[clip,width=9truecm]{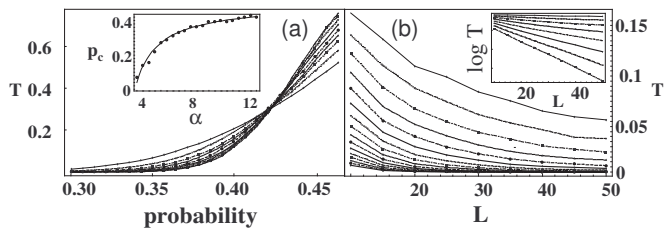}
\caption{\small Numerical results of the scattering matrix approach for $\alpha=12$ (see Eq.~\ref{TE}).
(a) The transmission coefficient $T$ as a function of Fermi energy for system
lengths $L=10,\ldots,50$.
Inset: The dependence of the critical point on
the parameter $\alpha$ . Bold points are raw data and the solid line
is the analytic curve (see text).
(b) The transmission coefficient $T$ as a function of $L$
for different concentrations, $p=0.41,0.4,\ldots,0.3$.
Inset: the same data on a semi-log plot demonstrating an $\exp(-L/\ksi(\e_F))$ dependence of the transmission.}

\vskip -0.5 truecm

\end{figure}
\begin{figure}[!h]
\centering
\includegraphics[clip,width=9truecm]{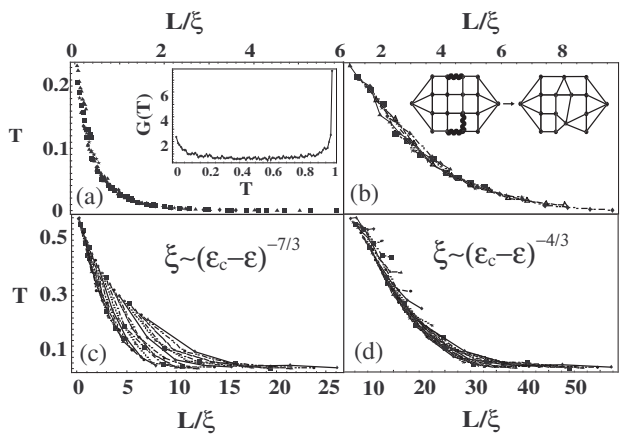}
\vskip -0.2 truecm \caption{\small (a) The raw data curves for $T$
(Fig.~3) are collapsed onto a single curve by scaling each curve
by an energy dependent localization length. This localization
length diverges at the critical energy with an exponent
$2.34\pm0.10$. The fixed distribution of $T$ at the critical point
is depicted in the inset. (b) Scaling of the data obtained by the
tight-binding approach. For energies $\epsilon=-0.050,-0.055,\ldots,-0.1$ the data collapse yields the exponent
$\nu=2.43 \pm 0.1$.
 Inset: the tight-binding approach -- sites connected by perfect links are merged and then the transmission
coefficient of the modified lattice is calculated using the
tight-binding hamiltonian (\ref{tb}). (c) Data collapse with the
same exponent as in (b), including a larger range of energies
$\epsilon=-0.01,-0.02,\ldots,-0.2$. This data can be better fitted
with an exponent $\nu=4/3$ (d). } \vskip -0.4 truecm
 \end{figure}
The exponent $\nu$ is found to be independent of the value of
$\alpha$ in (\ref{TE}). The critical probability, on the other
hand, does depend on $\alpha$ (inset of Fig.~3(a)). This
dependence can be understood as follows: The transmission through the
system at the critical point, $T_0$, is independent of length and
of the value of $\alpha$ and is given by an average over the fixed
distribution (inset of Fig.~4(a)). As the transmission rises with
Fermi energy $\e_F$, the critical Fermi energy for a given
$\alpha$, $\e_c(\alpha)$, is determined by the value of $\e_F$ at
which the transmission becomes equal to $T_0$.
 Since the distribution of link transmissions is exponential, the
conductance of the whole network is determined by the
critical transmission, the smallest transmission value such that the
links with better transmission percolate through the system
\cite{ambegaokar}. This transmission is given by
$\exp[\alpha(\e_c(\alpha)-\e_c)]$, with $\e_c=0$, the classical
critical Fermi energy. Equating this transmission to $T_0$, one
immediately finds $\e_c(\alpha)=\log(T_0)/\alpha$, or
$p_c(\alpha)=\half+\log(T_0)/\alpha$. This function is plotted in
the inset of Fig.~3(a) (solid curve), with $T_0=0.4$, and demonstrates an
excellent fit to the data. The average value of $\log(T)$ over the
fixed distribution found above gives $T_0=0.32$, in rough
agreement with the argument.
\par In order to study a model where the critical point is independent
of parameters, we develop a second approach, employing
a tight-binding Hamiltonian

\beq {\cal H} = \sum_{<i,j>} v_{ij}c^{\dagger }_{i} c_{j}+{
\mathrm H.c.} \quad. \label{tb}
\end{equation}

Determination of the parameters $v_{ij}$ follows the procedure
described above for determining the scattering matrix parameters.
We associate a random saddle-point energy $\e_{ij}$ with each link
joining the lattice sites $i$ and $j$. When $\e_{ij}>\e_F$  the hopping 
matrix element $v_{ij}$ for
that link is chosen such that it will give the transmission
coefficient (\ref{TE}) . When $\e_{ij}<\e_F$ the link is considered 
perfect, or SC. Since no value
of the hopping matrix element corresponds to perfect transmission
in the tight-binding approach, then when two sites are joined by
such a SC link, they are merged into a single site. This
guarantees that only when the SC links percolate, the conductance of
the system will not decay with length, so that $p_c$  will be
 independent of $\alpha$. The conductance is now
calculated numerically, by attaching a single-channel wire to both
sides of the lattice, and calculating the transmission through the
system for an electron at the Fermi energy. This calculation
involves two stages (see inset of Fig.~4(b)): (a) identifying
clusters of sites connected by SC links as a single site, and (b)
calculating $T$ by solving the tight-binding equations in the
reduced space \cite{meiraharony}. Data is generated for lattice
length $L=6,10,\ldots,34$, and averaged over $1500$ realizations.
Following the same scaling procedure as described above, the
curves collapse onto a
 single curve, a procedure that yields the critical exponent
 $\nu=2.43 \pm 0.1$ (Fig.~4(b)). We also calculate the conductance distribution at the
 critical point and find that it shows a two-peak structure, similar to that of
the S-matrix approach.
 Both the critical exponent and the critical point, $p_c=\half$
 are independent of the parameter $\alpha$, for a large range of
 values \cite{alpha=0}.
We also check the dependence of the critical
exponent $\nu$ for different functional forms of the tunneling probability $T(\e)$
  in addition to that described by (\ref{TE}).
Using, for example, the functional form obtained by Fertig and Halperin \cite{fertig}, yields very similar results,
 with a value
$\nu=2.49\pm0.1$. A similar value is obtained when the
link transmission coefficients are taken from a uniform distribution.
Thus, the model does not depend on the
exact functional form of $T(\e)$, or on its distribution.
The change in the critical exponent from its classical counterpart can be traced
 to the fact that as the Fermi energy, or concentration of SC links, change, the
 transmission amplitude through the quantum insulator is modified, as the bottle-neck link
 changes. We also note that repeating the same calculations without random
 phases does not yield any reasonable scaling of the data. Thus interference
 effects are crucial in obtaining the correct critical behavior.\par \noindent


We also note that if one tries to employ the scaling analysis
described above including energies that are not too close to the
critical one, the collapse of the data becomes worse and in fact
can be better fitted with the classical exponent $\nu=4/3$
(Fig.~4(c) and (d)). This is consistent with the above picture as
the difference between the classical and the quantum exponents
arise due to the additional dependence of the tunneling amplitude
on $\e-\e_c$, a dependence that becomes less relevant away from
the critical point. This observation may also explain why some
experiments report a classical percolation exponent for the QH transition
\cite{shahar}.


Since this model is practically identical to a model describing
the SIT, it is not surprising that the observed critical exponents
(on the insulating side) for the latter are very similar to those
observed for the former. However, unlike the
 QH transition, the SIT is not symmetric around the critical point, which is
 closer to the model described here. In fact, in the superconducting phase
 the mechanism described above for the change in the critical exponent from its classical percolation value
 due to the change in the transport in the insulating phase does not
 have any effect. This is consistent with the experimental observations of
 a critical exponent close to $\nu \approx 1.4$, even for samples
 where a larger exponent was observed on the insulating side \cite{sc2.8}.
 In addition, this relation between these two apparently different
 transitions also explains the fact that the critical resistance
 observed at the SIT seems to be distributed around values somewhat
 larger than $h/4e^2$, expected from duality\cite{orr}. This relation can be further
 investigated experimentally by studying the full resistance distribution
 at the SI threshold, similarly to such experiments for the QH transition\cite{cobden}.
 The similarity of the QH transition and the SIT has already been pointed out
 in \cite{shimshoni}, who emphasized the non-critical
 behavior and the role of decoherence. In fact,  the model introduced in this
 work allows the introduction of dephasing in a
straightforward way, by attaching current-conserving, phase-breaking reservoirs to
some fraction of the links \cite{buttiker}. The interplay between dephasing
and classical-quantum crossover will be explored in a future
communication \cite{yoni}.

We acknowledge fruitful discussions with A. Aharony
and O. Entin-Wohlman. This research has been funded by the ISF.


\begin{thebibliography}{99}
\bibitem{alpha=0}
Near $\alpha=0$, the data collapse is worse. At $\alpha=0$ no scaling
 is observed and a critical exponent cannot be determined.
\end{thebibliography}
\end{document}